\documentstyle [12pt] {article}

\catcode`@=12
\input epsf.tex
\def\DESepsf(#1 width #2){\epsfxsize=#2 \epsfbox{#1}}

\begin{document}
\thispagestyle{empty}
\begin{flushright} UCRHEP-T176\\May 1997\ \end{flushright}
\vspace{0.5in}
\begin{center} {\Large \bf Generic Consequences of a Supersymmetric\\} 
\vspace{0.1in} {\Large \bf U(1) Gauge Factor at the TeV Scale\\} 
\vspace{1.0in} {\bf E. Keith and Ernest Ma\\}
\vspace{0.1in} {\sl Department of Physics, University of California, 
Riverside, CA 92521, USA\\}
\vspace{1.0in}
\end{center}
\begin{abstract}\ We consider an arbitrary supersymmetric U(1) gauge factor 
at the TeV scale, under which the two Higgs superfields $H_{1,2}$ of the 
standard model are nontrivial. We assume that there is a singlet superfield 
$S$ such that $H_1 H_2 S$ is an allowed term in the superpotential. We 
discuss first the generic consequences of this hypothesis on the structure 
of the two-doublet Higgs sector at the electroweak energy scale, as well 
as $Z-Z'$ mixing and the neutralino sector. We then assume the existence 
of a grand unified symmetry and universal soft supersymmetry breaking terms 
at that scale. We further assume that the additional U(1) is broken 
radiatively via a superpotential term of the form $h h^c S$, where $h$ and 
$h^c$ are exotic color-triplet fields which appear in $E_6$ models.   We 
show that the U(1) breaking scale and the parameter $\tan \beta \equiv 
v_2/v_1$ are then both predicted as functions of the $H_1 H_2 S$ coupling.
\end{abstract}

\newpage
\baselineskip 24pt
\section{Introduction}

If supersymmetry is broken at the TeV energy scale and the standard $SU(3)_C 
\times SU(2)_L \times U(1)_Y$ gauge symmetry is not extended at that same 
scale, then the existence of supersymmetry above that scale protects the 
theory from nondecoupling contributions of physics above that scale. 
In other words, we get the Minimal Supersymmetric Standard Model (MSSM). 
However, if the gauge symmetry is extended also at the TeV energy scale 
and it breaks down to that of the standard model together with the 
supersymmetry, there will be in general new important phenomenological 
consequences at the TeV scale as well as the 100 GeV scale. With further 
simplifying assumptions, the parameters of the two scales may
also be related.

A particularly interesting extension of the MSSM is the inclusion of an 
extra U(1) factor at the TeV energy scale. The motivation for this could 
be theoretical. If the standard model is embedded in a larger symmetry 
group of rank greater than 4, such as $SO(10)$ [rank 5] or $E_6$ 
[rank 6], then an extra U(1) gauge factor is very possible at the TeV 
energy scale. This is especially true for the supersymmetric $E_6$ 
model\cite{1,2} based on the $E_8 \times E_8$ heterotic string. In 
particular, if only flux loops are invoked to break $E_6$ down to 
$SU(3)_C \times SU(2)_L \times U(1)_Y$, then a specific extra U(1) 
[conventionally known as $U(1)_\eta$] is obtained. Remarkably, $U(1)_\eta$ 
is also phenomenologically implicated\cite{3} by the experimental $R_b 
\equiv \Gamma (Z \rightarrow b \bar b)/\Gamma (Z \rightarrow hadrons)$ 
excess. Another possible clue is the totality of neutrino-oscillation
experiments (solar, atmospheric, and laboratory) which suggest that there 
are at least 4 neutrinos. This has been shown\cite{4} to have a natural 
explanation in terms of the $E_6$ superstring model with a specific U(1) 
called $U(1)_N$. 

In Sec.~2, we consider a generic extra supersymmetric U(1) gauge factor at 
the TeV energy scale with two doublet superfields $H_{1,2}$ and a singlet 
superfield $S$ such that $H_1 H_2 S$ is an allowed term in the superpotential. 
(Note that if $S$ has a nonzero charge under the aditional U(1), as is the 
case if the scalar component of $S$ is to acquire a nonzero vacuum 
expectation value (VEV) so as to break this U(1), then above this breaking 
scale, no $\mu H_1 H_2$ superpotential term exists.) We then derive its 
nondecoupling effects on the two-doublet Higgs sector at the 100 GeV scale.
In Sec.~3, we specialize to a class of $U(1)_\alpha$ models derivable from 
$E_6$, of which the $U(1)_\eta$ and $U(1)_N$ models are special cases. 
In Sec.~4, we discuss how the new $Z'$ mixes with the standard $Z$ in the 
general case, and formulate the effects in terms of the oblique parameters 
$\epsilon_{1,2,3}$ or $S,T,U$ in the $U(1)_\alpha$ models. We also discuss 
the generic neutralino sector. In Sec.~5, we show how supersymmetric scalar 
masses are affected by the extra D-terms from $U(1)_\alpha$. Combining this 
with the results of Sec.~2 and Sec.~3, and assuming universal soft 
supersymmetry breaking terms at the grand-unification scale, we show that 
there is a relationship between the $U(1)_\alpha$ vacuum expectation value 
and the well-known parameter $\tan \beta \equiv v_2/v_1$ used in the MSSM. 
Finally in Sec.~6, we have some concluding remarks.

\section{Tree-Level Nondecoupling at the 100 GeV Scale} 

As the U(1) gauge factor is broken together with the supersymmetry at the 
TeV scale, the resulting heavy scalar particles have nondecoupling 
contributions to the interactions of the light scalar particles.\cite{5} 
Consequently, the two-doublet Higgs structure is of a more general form 
than that of the Minimal Supersymmetric Standard Model (MSSM). Previous 
specific examples have been given.\cite{6,7,8} Here we present the most 
general analysis. We denote the scalar components of $H_1$, $H_2$, and $S$ 
as $\tilde \Phi_1$, $\Phi_2$, and $\chi$ respectively. Under $SU(3)_C \times 
SU(2)_L \times U(1)_Y \times U(1)_X$, we then have
\begin{eqnarray}
\tilde \Phi_1 &=& \left( \begin{array} {c} \bar \phi_1^0 \\ -\phi_1^- 
\end{array} \right) \sim (1,2,-{1 \over 2}; -a), \\ \Phi_2 &=& \left( 
\begin{array} {c} \phi_2^+ \\ \phi_2^0 \end{array} \right) \sim (1,2, 
{1 \over 2}; -1+a), \\ \chi &=& \chi^0 \sim (1,1,0;1), 
\end{eqnarray} 
where each last entry is the arbitrary assignment of that scalar multiplet
under the extra $U(1)_X$ with coupling $g_x$, assuming of course that the 
superpotential has the term $f H_1 H_2 S$. The corresponding scalar potential 
contains thus 
\begin{equation} 
V_F = f^2 [(\Phi_1^\dagger \Phi_2)(\Phi_2^\dagger \Phi_1) + (\Phi_1^\dagger 
\Phi_1 + \Phi_2^\dagger \Phi_2) \bar \chi \chi], 
\end{equation} 
and from the gauge interactions,
\begin{eqnarray} 
V_D &=& {1 \over 8} g_2^2 [(\Phi_1^\dagger \Phi_1)^2 + (\Phi_2^\dagger
\Phi_2)^2 + 2(\Phi_1^\dagger \Phi_1)(\Phi_2^\dagger \Phi_2) - 4(\Phi_1^\dagger
\Phi_2)(\Phi_2^\dagger \Phi_1)] \nonumber \\ &+& {1 \over 8} g_1^2 
[-\Phi_1^\dagger \Phi_1 + \Phi_2^\dagger \Phi_2]^2 \nonumber \\ &+& 
{1 \over 2} g_x^2 [-a \Phi_1^\dagger \Phi_1 - (1-a) \Phi_2^\dagger \Phi_2 
+ \bar \chi \chi]^2. 
\end{eqnarray} 
Let $\langle \chi \rangle = u$, then $\sqrt 2 Re \chi$ is a physical scalar 
boson with $m^2 = 2 g_x^2 u^2$, and the $(\Phi_1^\dagger \Phi_1) \sqrt 2 
Re \chi$ coupling is $\sqrt 2 u (f^2 -g_x^2 a)$. Hence the effective 
$(\Phi_1^\dagger \Phi_1)^2$ coupling $\lambda_1$ is given by 
\begin{eqnarray}
\lambda_1 &=& {1 \over 4} (g_1^2 + g_2^2) + g_x^2 a^2 - {{2(f^2 - g_x^2 a)^2} 
\over {2 g_x^2}} \nonumber \\ &=& {1 \over 4} (g_1^2 + g_2^2) + 2 a f^2 - 
{f^4 \over g_x^2}.
\end{eqnarray} 
Similarly,
\begin{eqnarray}
\lambda_2 &=& {1 \over 4} (g_1^2 + g_2^2) + 2 (1 - a) f^2 - {f^4 \over g_x^2}, 
\\ \lambda_3 &=& -{1 \over 4} g_1^2 + {1 \over 4} g_2^2 + f^2 - {f^4 \over 
g_x^2}, \\ \lambda_4 &=& -{1 \over 2} g_2^2 + f^2, 
\end{eqnarray} 
where the two-doublet Higgs potential has the generic form 
\begin{eqnarray} 
V &=& m_1^2 \Phi_1^\dagger \Phi_1 + m_2^2 \Phi_2^\dagger \Phi_2 +
m_{12}^2 (\Phi_1^\dagger \Phi_2 + \Phi_2^\dagger \Phi_1) + {1 \over 2} 
\lambda_1 (\Phi_1^\dagger \Phi_1)^2 \nonumber \\ &~& + {1 \over 2} 
\lambda_2 (\Phi_2^\dagger \Phi_2)^2 + \lambda_3 (\Phi_1^\dagger \Phi_1)
(\Phi_2^\dagger \Phi_2) + \lambda_4 (\Phi_1^\dagger \Phi_2)(\Phi_2^\dagger 
\Phi_1). 
\end{eqnarray} 
From Eqs.~(6) to (9), it is clear that the MSSM is recovered in the limit 
$f = 0$.  [Note that $m_{12}^2 \neq 0$ only after U(1) symmetry breaking and 
it would be proportional to $f$ if universal soft supersymmetry breaking is 
assumed.]  Let $\langle \phi_{1,2}^0 \rangle \equiv v_{1,2}$, $\tan
\beta \equiv v_2/v_1$, and $v^2 \equiv v_1^2 + v_2^2$, then this $V$ has 
an upper bound on the lighter of the two neutral scalar bosons given by 
\begin{equation} 
(m_h^2)_{max} = 2 v^2 [\lambda_1 \cos ^4 \beta + \lambda_2 \sin^4 \beta + 
2(\lambda_3 + \lambda_4) \sin^2 \beta \cos^2 \beta] + \epsilon, 
\end{equation} 
where we have added the radiative correction\cite{9} due to the $t$ quark 
and its supersymmetric scalar partners, {\it i.e.} 
\begin{equation}
\epsilon = {{3 g_2^2 m_t^4} \over {8 \pi^2 M_W^2}} \ln \left( 1 + {\tilde m^2 
\over m_t^2} \right).
\end{equation} 
We note also that the soft supersymmetry-breaking term $f A_f \Phi_1^\dagger 
\Phi_2 \chi + h.c.$ (from which we obtain $m_{12}^2 = f A_f u$) also 
contributes to $\lambda_4$ and generates some additional quartic scalar 
couplings.  However, we assume here that $f A_f/g_x^2 u$ is small, because 
we are mainly interested in the case where the elctroweak Higgs sector 
has two relatively light doublets and not just one light doublet. 
Using Eqs.~(6) to (9), we obtain
\begin{equation} 
(m_h^2)_{max} = M_Z^2 \cos^2 2 \beta + \epsilon + {f^2 \over {\sqrt 2 G_F}}
\left[ A - {f^2 \over g_x^2} \right],
\end{equation} 
where
\begin{equation} 
A = {3 \over 2} + (2 a - 1) \cos 2 \beta - {1 \over 2} \cos^2 2 \beta.
\end{equation} 
If $A > 0$, then the MSSM bound can be exceeded. However, $f^2$ is still
constrained from the requirement that $V$ be bounded from below.  We note 
here that although $V_F$ of Eq.~(4) and $V_D$ of Eq.~(5) are nonnegative 
for any value of $f$, $V$ of Eq.~(10) is not automatically bounded from 
below.  This simply means that if $f$ is too large, the minimum of the 
original potential only breaks the extra U(1) but not the electroweak 
gauge symmetry.  Given $g_x$ 
and $a$, we can vary $\cos 2 \beta$ and $f$ to find the largest numerical 
value of $m_h$. We show in Fig.~1 this upper bound on $m_h$ as a function of 
$g_x^2$ for several specific values of $a$. The value $a_0$ is chosen in the 
top curve to maximize $m_h$ for a given value of $g_x^2$. This
upper bound increases as $g_x^2$ increases. However, it is reasonable to 
assume that $g_x$ cannot be too large. In fact, in the specific models to be 
discussed in the next section, $g_x^2 < 0.16$. As shown in Fig.~1, even for 
$g_x^2 = 0.5$, the upper bound is only about 190 GeV. 

It should be mentioned that an upper bound on $m_h$ has been previously 
obtained\cite{10} assuming that there is no extra U(1) at the 
supersymmetry-breaking scale. However, the same proof also goes through 
with an extra U(1). We improve on Ref.[10] in this case by computing
exactly how the off-diagonal nondecoupling terms affect the upper bound 
on $m_h$, resulting in Fig.~1 as shown.  If $f A_f/g_x^2 u$ is not small 
as we have assumed, then the reduction to $V$ of Eq.~(10) is not valid. 
In this case, the electroweak Higgs sector consists of effectively only 
one Higgs doublet with $m_h$ bounded by a function which involves also 
$A_f$.\cite{11}

\section{U(1) Gauge Factor from E(6)}

As already mentioned in the Introduction, an extra supersymmetric U(1) gauge 
factor at the TeV scale is a very viable possibility from the spontaneous 
breakdown of $E_6$. Consider the following sequential reduction: 
\begin{eqnarray} 
E_6 &\rightarrow& SO(10) ~[\times U(1)_\psi] \\ SO(10) &\rightarrow& SU(5) 
~[\times U(1)_\chi] \\ SU(5) &\rightarrow& SU(3)_C \times
SU(2)_L ~[\times U(1)_Y]. 
\end{eqnarray} 
At each step, a U(1) gauge factor may or may not appear, depending on the 
details of the symmetry breaking. Assuming that a single extra U(1)
survives down to the TeV energy scale, then it is generally given by a linear 
combination of $U(1)_\psi$ and $U(1)_\chi$ which we will call $U(1)_\alpha$. 

Under the maximal subgroup $SU(3)_C \times SU(3)_L \times SU(3)_R$, the 
fundamental representation of $E_6$ is given by 
\begin{equation} 
{\bf 27} = (3,3,1) + (3^*,1,3^*) + (1,3^*,3). 
\end{equation} 
Under the subgroup $SU(5) \times U(1)_\psi \times U(1)_\chi$, we then have 
\begin{eqnarray} 
{\bf 27} &=& (10; 1, -1) ~[(u,d), u^c, e^c] \nonumber \\ &+& (5^*;
1, 3) ~[d^c, (\nu_e, e)] \nonumber \\ &+& (1; 1, -5) ~[N] \nonumber \\ &+& 
(5; -2, 2) ~[h, (E^c, N_E^c)] \nonumber \\ &+& (5^*; -2, -2) ~[h^c, 
(\nu_E, E)] \nonumber \\ &+& (1; 4, 0) ~[S], 
\end{eqnarray} 
where the U(1) charges refer to $2 \sqrt 6 Q_\psi$ and $2 \sqrt {10}
Q_\chi$. Note that the known quarks and leptons are contained in $(10; 1, -1)$ 
and $(5^*; 1, 3)$, and the two Higgs scalar doublets are represented by 
$(\nu_E, E)$ and $(E^c, N_E^c)$. Let
\begin{equation} 
Q_\alpha = Q_\psi \cos \alpha - Q_\chi \sin \alpha, 
\end{equation} 
then the $\eta$-model\cite{1,2} is obtained with $\tan \alpha = \sqrt {3/5}$ 
and we have
\begin{eqnarray} 
{\bf 27} &=& (10;2) + (5^*; -1) + (1;5) \nonumber \\ &+& (5; -4) + (5^*; -1)
+ (1;5),
\end{eqnarray} 
where $2 \sqrt {15} Q_\eta$ is denoted; and the $N$-model\cite{4} is obtained
with $\tan \alpha = -1/\sqrt {15}$ resulting in 
\begin{eqnarray} 
{\bf 27} &=& (10;1) + (5^*;2) + (1;0) \nonumber \\ &+& (5;-2) + (5^*;-3) 
+ (1;5),
\end{eqnarray} 
where $2 \sqrt {10} Q_N$ is denoted. This model is so called because the
superfield $N$ has $Q_N = 0$. It allows $S$ to be a naturally light singlet 
neutrino and is ideally suited to explain the totality of all 
neutrino-oscillation experiments, {\it i.e.} solar\cite{12}, 
atmospheric\cite{13}, and laboratory\cite{14}. It is also a natural
consequence of an alternative $SO(10)$ decomposition\cite{15} of $E_6$, 
{\it i.e.}
\begin{eqnarray} 
{\bf 16} &=& [(u,d), u^c, e^c; h^c, (\nu_E,E); S], \\ {\bf 10} &=& [h, (E^c,
N_E^c); d^c, (\nu_e, e)], \\ {\bf 1} &=& [N],
\end{eqnarray} 
which differs from the conventional assignment by how the $SU(5)$ multiplets
are embedded.

Identifying $\tilde \Phi_1$, $\Phi_2$, and $\chi$ with the scalar components 
of $(\nu_E, E)$, $(E^c, N_E^c)$, and $S$ of which we can choose one copy of 
each via a discrete symmetry\cite{4} to be the ones with VEVs, we see that the 
general anaylsis of the previous section is  applicable for this class
of U(1)-extended models. (Of course, more than one copy of $(\nu_E, E)$,
$(E^c, N_E^c)$, or $S$ could have VEVs but that would lead to a much less 
constrained scenario.) Assuming that $U(1)_\alpha$ is normalized in the 
same way as $U(1)_Y$, we find it to be a very good approximation\cite{8} 
to have $g_\alpha^2 = (5/3) g_1^2$. We then obtain for the $\eta$-model, 
\begin{equation} 
g_x^2 = {25 \over 36} g_1^2, ~~~ a = {1 \over 5}, 
\end{equation} 
and for the $N$-model,
\begin{equation} 
g_x^2 = {25 \over 24} g_1^2, ~~~ a = {3 \over 5}, 
\end{equation} 
whereas in the exotic left-right model\cite{6,14}, 
\begin{equation} 
g_x^2 = {{(g_1^2+g_2^2) (1-\sin^2 \theta_W)^2} \over {4 (1-2\sin^2 
\theta_W)}}, ~~~ a = \tan^2 \theta_W.
\end{equation} 
These three specific points have been singled out in Fig.~1. Furthermore, when
we take the squark masses to be about 1 TeV we find the largest numerical 
value of $m_h$ in the $U(1)_\alpha$ models to be about 142 GeV , as compared 
to 128 GeV in the MSSM, and it is achieved with
\begin{equation}
\tan \alpha = - {{2 \sqrt {3/5} \cos 2 \beta} \over {3 - \cos^2 2 \beta}}, 
\end{equation}
which is possible in the $\eta$-model, {\it i.e.} $\tan\alpha = \sqrt {3/5}$ 
and $\cos 2 \beta = -1$.

\section{Z - Z' and Neutralino Sectors}

The part of the Lagrangian containing the interaction of $\Phi_{1,2}$ and 
$\chi$ with the vector gauge bosons $A_i (i = 1,2,3)$, $B$, and $Z'$ 
belonging to the gauge factors $SU(2)_L$, $U(1)_Y$, and $U(1)_X$ respectively 
is given by
\begin{eqnarray} 
{\cal L} &=& |(\partial^\mu - {{i g_2} \over 2} \tau_i A_i^\mu + {{i g_1}
\over 2} B^\mu + i g_x a Z'^\mu) \tilde \Phi_1|^2 \nonumber \\ &+& 
|(\partial^\mu - {{i g_2} \over 2} \tau_i A_i^\mu - {{i g_1} \over 2} B^\mu 
+ i g_x (1-a) Z'^\mu) \Phi_2|^2 \nonumber \\ &+& |(\partial^\mu - 
i g_x Z'^\mu) \chi|^2, 
\end{eqnarray} 
where $\tau_i$ are the usual $2 \times 2$ Pauli matrices. With the definition 
$Z \equiv (g_2 A_3 - g_1 B)/g_Z$, where $g_Z \equiv \sqrt {g_1^2 + g_2^2}$, 
the mass-squared matrix spanning $Z$ and $Z'$ is given by
\begin{equation} 
{\cal M}^2_{Z,Z'} = \left[ \begin{array} {c@{\quad}c} (1/2)g_Z^2 (v_1^2 +
v_2^2) & g_Z g_x [-a v_1^2 + (1-a) v_2^2] \\ g_Z g_x [-a v_1^2 + (1-a) v_2^2] 
& 2 g_x^2 [u^2 + a^2 v_1^2 + (1-a)^2 v_2^2] \end{array} \right]. 
\end{equation} 
Let the mass eigenstates of the $Z-Z'$ system be 
\begin{equation} 
Z_1 = Z \cos \theta + Z' \sin \theta, ~~~ Z_2 = -Z \sin \theta + Z' 
\cos \theta,
\end{equation} 
then the experimentally observed neutral gauge boson is identified in this
model as $Z_1$, with mass given by
\begin{equation} 
M^2_{Z_1} \equiv M_Z^2 \simeq {1 \over 2} g_Z^2 v^2 \left[ 1 - (\sin^2 \beta
- a)^2 {v^2 \over u^2} \right],
\end{equation} 
and
\begin{equation}
\theta \simeq - {g_Z \over {2 g_x}} (\sin^2 \beta - a) {v^2 \over u^2}. 
\end{equation} 
Note that $Z_2$ has essentially the same mass as the physical scalar boson 
$\sqrt 2 Re \chi$ discussed earlier. 

So far, our discussion of the $Z-Z'$ sector is completely general. However, 
in order to make contact with experiment, we have to specify how $Z'$ 
interacts with the known quarks and leptons. In the class of $U(1)_\alpha$ 
models from $E_6$, all such couplings are determined. In particular, we have
\begin{equation} 
g_x = \sqrt {2 \over 3} g_\alpha \cos \alpha , ~~~ a = {1 \over 2} \left( 1 -
\sqrt {3 \over 5} \tan \alpha \right). 
\end{equation} 
Using the leptonic widths and
forward-backward asymmetries of $Z_1$ decay, the deviations from the standard 
model are conveniently parametrized\cite{16}: 
\begin{eqnarray}
\epsilon_1 &=& \left[ \sin^4 \beta - {1 \over 4} \left( 1 - \sqrt {3 \over 5} 
\tan \alpha \right)^2 \right] {v^2 \over u^2} ~\simeq~ \alpha T, \\ \epsilon_2 
&=& {1 \over 4} (3 - \sqrt {15} \tan \alpha) \left[ \sin^2 \beta - {1 \over 2} 
\left( 1 - \sqrt {3 \over 5} \tan \alpha \right) \right] {v^2 \over u^2} 
~\simeq~ - {{\alpha U} \over {4 \sin^2 \theta_W}}, \\ \epsilon_3 &=& 
{1 \over 4} \left[ 1 - 3 \sqrt {3 \over 5} \tan \alpha + {1 \over {2 \sin^2
\theta_W}} \left( 1 + \sqrt {3 \over 5} \tan \alpha \right) \right] \nonumber 
\\ &\times& \left[ \sin^2 \beta - {1 \over 2} \left( 1 - \sqrt {3 \over 5} 
\tan \alpha \right) \right] {v^2 \over u^2} ~\simeq~ {{\alpha S} \over 
{4 \sin^2 \theta_W}}. 
\end{eqnarray} 
Since the experimental errors on these quantities are fractions of a percent, 
$u \sim$ TeV is allowed.

In the MSSM, there are four neutralinos (two gauge fermions and two Higgs 
fermions) which mix in a well-known $4 \times 4$ mass matrix\cite{17}. 
Here we have six neutralinos: the gauginos of $U(1)_Y$ and the third 
component of $SU(2)_L$, the Higgsinos of $\bar \phi_1^0$ and $\phi_2^0$, 
the $U(1)_X$ gaugino and the $\chi$ Higgsino. The corresponding mass matrix is
then given by
\begin{equation} 
{\cal M_N} = \left[ \begin{array}
{c@{\quad}c@{\quad}c@{\quad}c@{\quad}c@{\quad}c} 
M_1 & 0 & -g_1 v_1/\sqrt 2 & g_1 v_2/\sqrt 2 & 0 & 0 \\ 
0 & M_2 & g_2 v_1/\sqrt 2 & -g_2 v_2/\sqrt 2 & 0 & 0 \\ 
-g_1 v_1/\sqrt 2 & g_2 v_1/\sqrt 2 & 0 & fu & -g_x a v_1 \sqrt 2 & f v_2 \\
 g_1 v_2/\sqrt 2 & -g_2 v_2/\sqrt 2 & fu & 0 & -g_x (1-a) v_2 \sqrt 2 & f v_1 
\\ 
0 & 0 & -g_x a v_1 \sqrt 2 & -g_x (1-a) v_2 \sqrt 2 & M_x & g_x u \sqrt 2\\ 
0 & 0 & f v_2 & f v_1 & g_x u \sqrt 2 & 0 
\end{array} \right], 
\end{equation} 
where $M_{1,x,2}$ are allowed U(1) and SU(2) gauge-invariant Majorana mass 
terms which break the supersymmetry softly. Note that without the last two 
rows and columns, the above matrix does reduce to that of the MSSM if $fu$ 
is identified with $-\mu$. However, the $\mu$ parameter in the MSSM is 
unconstrained, whereas here $fu$ is bounded and $f$ itself appears in the Higgs
potential.

Since $g_x u$ should be of order TeV, the neutralino mass matrix $\cal M_N$ 
reduces to either a $4 \times 4$ or $2 \times 2$ matrix, depending on whether 
$f$ is much less than $g_x$ or not. In the former case, it reduces to that of 
the MSSM but with the stipulation that the $\mu$ parameter must be small, 
{\it i.e.} of order 100 GeV. This means that the two gauginos mix 
significantly with the two Higgsinos and the lightest supersymmetric particle 
(LSP) is likely to have nonnegligible components from all four states. In the 
latter case, the effective $2 \times 2$ mass matrix becomes
\begin{equation} 
{\cal M'_N} = \left[ \begin{array} {c@{\quad}c} M_1 + g_1^2 v_1 v_2/fu & -g_1
g_2 v_1 v_2/fu \\ -g_1 g_2 v_1 v_2/fu & M_2 + g_2^2 v_1 v_2/fu \end{array} 
\right].
\end{equation} 
Since $v_1 v_2/u$ is small, the mass eigenstates of $\cal M'_N$ are
approximately the gauginos $\tilde B$ and $\tilde W_3$, with masses $M_1$ 
and $M_2$ respectively. In supergravity models with uniform gaugino masses 
at the GUT breaking scale, 
\begin{equation} 
M_1 = {{5 g_1^2} \over {3 g_2^2}} M_2 \simeq 0.5 M_2, 
\end{equation} 
hence $\tilde B$ would be the LSP, which makes it a good
candidate for cold dark matter.

\section{Supersymmetric Scalar Masses}

The spontaneous breaking of the additional U(1) gauge factor at the TeV scale 
is not possible without also breaking the supersymmetry.\cite{4}  As a 
reasonable and predictive procedure, we will adopt the common hypothesis that 
soft supersymmetry breaking operators appear at the grand-unification (GUT) 
scale as the result of a hidden sector which is linked to the observable 
sector only through gravity.  Moreover these terms will be assumed to be 
universal, {\it i.e.} of the same magnitude for all fields.

Consider now the masses of the supersymmetric scalar partners of the quarks 
and leptons:
\begin{equation} 
m_B^2 = m_0^2 + m_R^2 + m_F^2 + m_D^2,
\end{equation} 
where $m_0$ is a universal soft supersymmetry breaking mass at the GUT scale, 
$m_R^2$ is a correction generated by the renormalization-group equations 
running from the GUT scale down to the TeV scale, $m_F$ is the explicit mass 
of the fermion partner, and $m_D^2$ is a term induced by gauge symmetry 
breaking with rank reduction and can be expressed in terms of the gauge-boson 
masses. In the MSSM, $m_D^2$ is of order $M_Z^2$ and does not change $m_B$ 
significantly. In the $U(1)_\alpha$-extended model, $m_D^2$ is of order 
$M_{Z'}^2$ and will affect $m_B$ in a nontrivial way. For example, in the
case of ordinary quarks and leptons,
\begin{eqnarray}
\Delta m_D^2 (10; 1, -1) &=& {1 \over 8} M_{Z'}^2 \left( 1 + \sqrt {3 \over 5} 
\tan \alpha \right), \\ \Delta m_D^2 (5^*; 1, 3) &=& {1 \over 8} M_{Z'}^2 
\left( 1 - 3 \sqrt {3 \over 5} \tan \alpha \right). 
\end{eqnarray} 
This would have important consequences on the experimental search of 
supersymmetric particles. In fact, if $m_F$ is not too large, it is 
possible for the exotic scalars (which may be interpreted as leptoquarks 
depending on their Yukawa couplings) to be lighter than the usual scalar 
quarks and leptons.  We have already discussed this issue in Ref.~[18]. 

Assuming Eq.~(42), we first consider the spontaneous breaking of 
$U(1)_\alpha$, {\it i.e.} $\langle \chi \rangle = u$, which requires 
$m_\chi^2$ to be negative. This may be achieved by considering the 
superpotential
\begin{equation} 
W = f H_1 H_2 S + f' h h^c S + \lambda_t H_2 Q_3 t^c,
\end{equation} 
(where we have omitted the rest of the MSSM Yukawa couplings) together with the
trilinear soft supersymmetry-breaking terms
\begin{equation} 
V_{soft} = f A_f \Phi_1^\dagger \Phi_2 \chi + f' A_{f'} \tilde h \tilde h^c
\chi + \lambda_t A_t \Phi_2 {\tilde Q}_3 {\tilde t}^c\, 
\end{equation} 
along with the soft supersymmetry-breaking scalar masses. Starting with a 
wide range of given values of $m_0$, the universal gaugino mass
$m_{1/2}$, and the universal trilinear massive parameter $A_0$ at the GUT 
scale, we find that $m_\chi^2$ does indeed turn negative near the TeV energy 
scale for many typical values of $f$ and $f'$. An example of this is given 
in Fig. 2. The evolution of $m_\chi^2$ is mostly driven by $f'$, but
$f$ also contributes primarily through its direct effect on $A_{f'}$. 
From the negative value of $m_\chi^2$ at the TeV scale, we then obtain 
the predicted mass of $Z'$, {\it i.e.} $M_{Z'} = (-2 m_\chi^2)^{1/2}$,  
which is also the mass of the physical scalar boson $\sqrt 2 Re \chi$.
However, as we will discuss shortly, the mass of the $Z'$ so obtained 
must also be consistent with the
desired electroweak symmetry breaking conditions. 

We assume that the top quark's pole mass is 175 GeV, and that at 1 TeV 
$\alpha_s = 0.1$ which corresponds to $\alpha_s (M_Z)\approx 0.12$. 
We will also  assume that at the TeV scale and above, the particle content 
of the model is that of three complete {\bf 27}'s of $E_6$ and some additional 
field content so as to achieve gauge coupling unification. The
additional field content could be near the unification scale and hence provide 
threshold corrections that allow the gauge couplings to unify, perhaps even at 
the string compactification scale. Another possibility is to add an 
anomaly-free pair of $SU(2)_L$ doublet fields so as to mimic gauge coupling 
unification in the MSSM. Such an example is discussed for the $\alpha =N$ 
model of Ref. [8]. This model has the same unification scale as
is possible in the MSSM.  In calculating the gauge-coupling beta functions, 
we will in fact assume the field content of that model, but the choice of 
additional matter fields or threshold corrections to bring about gauge 
coupling unification has no significant effect on our calculation. 
The fact that such models have three complete {\bf 27}'s has the noteworthy 
implication that the gauge coupling at the unification scale is
approximately the strong coupling.  The reason is that with three copies of 
$h$ and $h^c$, the beta function for $\alpha_s$ is zero in one loop above 
the TeV scale.  Similarly, the gluino mass also does not evolve in this 
approximation.

Defining 
\begin{eqnarray}
{\cal D}\equiv 8\pi^2 {d\over d\ln{\mu}} 
\end{eqnarray}
(where $\mu$ is the scale), the relevant renormalization group equations are:
\begin{eqnarray}
{\cal D}\ln{\lambda_t^2} &=& -\sum_i{c_i^{(t)}}g_i^2 + 6 \lambda_t^2 + f^2\, ,\\
{\cal D}\ln{f^2} &=& -\sum_i{c_i^{(f)}}g_i^2 + 3 \lambda_t^2 + 4 f^2 + 3 {f'}^2\, ,\\
{\cal D}\ln{{f'}^2} &=& -\sum_i{c_i^{(f')}}g_i^2 + 3 f^2 + 5 {f'}^2\, ,
\end{eqnarray}
for the Yukawa couplings, 
\begin{eqnarray}
{\cal D}A_t &=& \sum_i{c_i^{(t)}}g_i^2 M_i + 6 \lambda_t^2 A_t + f^2 A_f\, ,\\
{\cal D}A_f &=& \sum_i{c_i^{(f)}}g_i^2 M_i + 3 \lambda_t^2 A_t 
+ 4 f^2 A_f + 3 {f'}^2 A_{f'}\, ,\\
{\cal D}A_{f'} &=& \sum_i{c_i^{(f')}}g_i^2 M_i + 3 f^2 A_f + 5 {f'}^2 A_{f'}\, ,
\end{eqnarray}
for the trilinear scalar parameters $A_i$, and 
\begin{eqnarray}
{\cal D}m_S^2 &=& -\sum_i{c_i^{(S)}}g_i^2 + 2 f^2 X_f + 3 {f'}^2 X_{f'}\, ,\\
{\cal D}m_h^2 &=& -\sum_i{c_i^{(h)}}g_i^2 + {f'}^2 X_{f'}\, ,\\
{\cal D}m_{h^c}^2 &=& -\sum_i{c_i^{(h^c )}}g_i^2 + {f'}^2 X_{f'}\, ,\\
{\cal D}m_{\Phi_1}^2 &=& -\sum_i{c_i^{(\Phi_1 )}}g_i^2 + {f}^2 X_{f}\, ,\\
{\cal D}m_{\Phi_2}^2 &=& -\sum_i{c_i^{(\Phi_2 )}}g_i^2 + 3 \lambda_2^2 X_t  
+ {f}^2 X_{f}\, ,\\
{\cal D}m_{Q_3}^2 &=& -\sum_i{c_i^{(Q_3 )}}g_i^2 +\lambda_2^2 X_t\, ,\\ 
{\cal D}m_{t^c}^2 &=& -\sum_i{c_i^{(t^c )}}g_i^2 +\lambda_2^2 X_t\, , 
\end{eqnarray}
where we have defined
\begin{eqnarray}
X_t\equiv m_{Q_3}^2 + m_{t^c}^2 + m_{\Phi_2}^2 + A_t^2\, ,\\
X_f\equiv m_{S}^2 + m_{\Phi_1}^2 + m_{\Phi_2}^2 + A_f^2\, ,\\
X_{f'}\equiv m_{S}^2 + m_{h}^2 + m_{h^c}^2 + A_{f'}^2\, ,
\end{eqnarray}
and the coefficients $c_i^{({\rm field})}$ have the obvious values. 
Further, the gaugino mass $M_i$  scales the same as  $\alpha_i$. 
These equations are modified in an obvious manner if $\tan\beta$ is large 
enough that $\lambda_b$ and $\lambda_\tau$ cannot be ignored or if there are 
more than one sizeable coupling serving the purpose of $f'$, which is 
certainly possible since we have three copies of $h$ and $h^c$ in these 
models.    

A very important outcome of Eq.~(42) is that the $U(1)_\alpha$ and electroweak 
symmetry breakings are related. To see this, go back to the two-doublet Higgs 
potential $V$ of Eq.~(10). Using Eqs.~(6) to (9) and Eq.~(35), we can express 
the parameters $m_{12}^2$, $m_1^2$, and $m_2^2$ in terms of the mass of the 
pseudoscalar boson $m_A$, and $\tan \beta$. 
\begin{eqnarray} 
m_{12}^2 &=& -m_A^2 \sin \beta \cos \beta, \\ m_1^2 &=& m_A^2 \sin^2 \beta
-{1 \over 2} M_Z^2 \cos 2 \beta \nonumber \\ &~& - {{2 f^2} \over g_Z^2} 
M_Z^2 \left[ 2 \sin^2 \beta + \left( 1 - \sqrt {3 \over 5} \tan \alpha 
\right) \cos^2 \beta - {{3 f^2} \over {2 \cos^2 \alpha ~g_\alpha^2}} \right], 
\\ m_2^2 &=& m_A^2 \cos^2 \beta + {1 \over 2} M_Z^2 \cos 2 \beta \nonumber \\ 
&~& -{{2 f^2} \over g_Z^2} M_Z^2 \left[ 2 \cos^2 \beta + \left( 1 + \sqrt
{3 \over 5} \tan \alpha \right) \sin^2 \beta - {{3 f^2} \over {2 \cos^2 
\alpha ~g_\alpha^2}} \right]. 
\end{eqnarray} 
On the other hand, using Eq.~(42), we have 
\begin{eqnarray} 
m_{12}^2 &=& f A_f u, \\ m_1^2 &=& m_0^2 + m_{R1}^2 + f^2 u^2 - {1 \over 4} 
\left( 1 - \sqrt {3 \over 5} \tan \alpha \right) M_{Z'}^2, \\ m_2^2 &=& m_0^2 
+ m_{R2}^2 + f^2 u^2 - {1 \over 4} \left( 1 + \sqrt {3 \over 5} \tan \alpha 
\right) M_{Z'}^2,
\end{eqnarray} 
where $m_{R1}^2$ and $m_{R2}^2$ differ in that $\lambda_t$ (the Yukawa coupling
of $\Phi_2$ to the $t$ quark) contributes to the latter but not to the former. 
Both depend on $m_0$, $m_{1/2}$, $A_0$, the various gauge couplings $g_i$, as 
well as $f$ and $f'$. Matching Eqs.~(64) to (66) with Eqs.~(67) to (69) 
allows us to determine $u$ and $\tan \beta$ for a
given set of parameters at the grand-unification scale. 

We will now briefly discuss our method for finding $u$ and $\tan \beta$ for a
given set of universal soft supersymmetry-breaking parameters $m_{\tilde g},
m_0,A_0$ at the GUT scale and the Yukawa coupling $f$, when such a solution 
exists. First, we guess a value for $\tan\beta$ so as to choose a value for 
$\lambda_t$. We then form a table $[M_{Z'},m_{R1}^2,m_{R2}^2,A_f](f')$ for 
many very closely spaced values of $f'$ extending up to where $f'(M_G)$ 
reaches its perturbative limit. By ``closely spaced values of $f'$," we mean 
that between two consecutive entries in the table, none of the four parameters 
differs by more than one percent. Second, we guess a value for
$M_{Z'}$ which lies within the range in the table, so as to choose 
$m_{R1}^2,m_{R2}^2,A_f$ from the entry of the table which has
$M_{Z'}$ closest to this value.   Third, we equate the right-hand sides 
of Eq. (65) $+$ Eq. (66) and Eq. (68) $+$ Eq. (69) to solve for $m_A^2$ 
as a linear function of $u^2$ and $\cos^2{\beta}$. Fourth,  using the 
previous result for $m_A^2$ we equate the right-hand sides  of Eq. (65) 
$-$ Eq. (66) and Eq. (68) $-$ Eq. (69) to solve for $u^2$ as a function of 
$\cos^2{\beta}$ of the form of a linear function divided by another linear 
function. Fifth, using the expressions from the previous two steps we 
equate the right-hand sides of Eq. (64) with that of Eq. (67) and
solve numerically for $\cos^2{\beta}$, and hence $\tan\beta$,  by first 
searching for a root close to the value corresponding to our original
guess for $\tan\beta$. In doing this fifth step, one needs to choose 
$f u >0$ or $f u<0$ analagous to $\mu >0$ or $\mu <0$ in the MSSM, and then 
check that the solution is consistent with 
$m_A^2=-fA_fu/\sin\beta\cos\beta >0$. In fact, taking all Yukawa couplings 
and $\tan\beta$ to be positive as well as our convention for the trilinear 
coupling parameters $A_i$, solutions exist only for $u<0$. Next, if a 
solution to these steps has been found, we start the  entire cycle  over 
using  the values for $\tan\beta$ and
$M_{Z'}$ just calculated as the new ``guessed" values. This iteration is
continued until the predicted $\tan\beta$ and
$M_{Z'}$ become fixed to a reasonable accuracy (we demand about five-percent 
accuracy). This process can be speeded up by adding a sixth step to the cycle 
which repeats the third through fifth steps until the prediction for  
$\tan\beta$ and
$M_{Z'}$ become fixed for the table found in the second step of the cycle.   

Before we discuss our results, we remind the reader that for the case that $A_f$
could be small
$f$ has a maximum  possible value that comes from requiring that the Higgs
potential be bounded  from below and which depends on the additional U(1). We
plot this maximum  value $f_{\rm max}$ as a function of $\alpha $ (see Eq. (20))
in Fig. 3.  In particular, the $\eta-$model requires $f$ to be less than about
0.35,  whereas for $\alpha =0$, $f$ could be as great as 0.46. Note that as 
$|\alpha |$ approaces $\pi /2$, $f_{\rm max}$ approaches 0.  From
Fig. 2(a), one can see that if $f$ is small enough so that
$\alpha =\eta$ is allowed, then $f (M_G)$ will always be perturbatively 
small for a perturbatively valued $f'$. In our examples,
we will only be interested in values of $f< 0.35$. 

In Fig. 4, we show the predicted values of $\tan\beta$ and
$M_{Z'}$ as a function of $\alpha$ for $f=0.345$ and 
$m_{\tilde g}=200$ GeV, $A_0=650$ GeV and
$m_0=650$ GeV. In accordance with Fig. 3, we are only interested in showing 
$|\alpha |$ less than about 0.7. We have also plotted the magnitude $|u|$ 
of the VEV of the singlet Higgs boson $\chi$ and the mass $f'|u|$ of 
the exotic fermion $h(h^c)$. In Fig. 5, we show the similar situation for  
$f=0.345$ and  $m_{\tilde g}=300$ GeV, $A_0=950$ GeV and
$m_0=950$ GeV. These two figures are quite similar except that the mass scale 
in Fig. 5(b) has been pushed up relative to that shown in Fig. 4(b).  
These choices of soft supersymmetry-breaking parameters are fairly typical 
in that generally we need $m_0$ to be at least twice as great as
$m_{\tilde g}$ to find a solution.  Further, if we want to have a solution 
for all $\alpha$ less than some value, $A_0$ must be positive and of order 
$m_0$.

In Figs. 6-9, we illustrate the effects of varying the parameters 
$f,m_{\tilde g},A_0,m_0$ for a fixed value of $\alpha =\eta$. We look at 
the solutions for $\tan\beta$ and $M_{Z'}$ (as well as $|u|$ and
$f'|u|$) when the four input parameters are varied one at a time around 
the point $f=0.345$ and  $m_{\tilde g}=250$ GeV, $A_0=650$ GeV and
$m_0=650$ GeV. Note from Fig. 6 that with decreasing $f$, $\tan\beta$ 
and $M_{Z'}$ both increase. We do not extend $f$ above 0.345 so as to avoid 
the upper bound coming from Fig. 3.  We find also that we cannot decrease 
$f$ much below 0.32 for this example and still have a 
solution for the electroweak breaking. To use smaller values of $f$, 
one would have to increase the scale of the soft supersymmetry-breaking 
parameters. In Fig. 7,  we look at the effect of varying $A_0$. The range of 
$A_0$ examined is restricted because any extension in either direction  
would require values of $M_{Z'}$ larger than can be reached via the 
$f'hh^cS$ term with $f'$ within the 
perturbative regime. In Fig. 8, we vary $m_0$.  Note that with increasing 
$m_0$, the predicted $\tan\beta$ increases significantly and $M_{Z'}$ 
decreases. In this example, increasing $m_0$ beyond 1200 GeV would predict 
an $M_{Z'}$ less than 500 GeV and a $\tan\beta$ greater than 10. 
The lower limit of 500 GeV for $m_0$ used here is due to the same reason
as just given for the range of $A_0$ plotted in the previous figure. 
In Fig. 9, we show the effect of varying the gluino mass which is also here 
the GUT scale universal gaugino mass. With increasing gluino
mass, $\tan\beta$ decreases while $M_{Z'}$ increases. The upper limit 
of 350 GeV used here for the gluino mass again corresponds to about the 
size of that parameter for this example where increasing it anymore would 
require values of $|u|$ larger than can be reached perturbatively 
through the renormalization-group equations. We find the general trends 
of Fig. 6-9 to be typical of other choices of paramter values where 
consistent solutions exist. 

If $m_0$ is demanded to be less than about 1 TeV, then in general 
$\tan\beta <10$, where the $b$ and $\tau$ Yukawa couplings are small 
enough not to contribute significantly to the renormalization-group 
equations.  It is interesting to note that in contrast to the MSSM, 
where $m_1^2 - m_2^2 = - m^2_{R_2}(\lambda_t) = -(m_A^2 + m_Z^2) \cos2\beta$,  
solutions with $\tan\beta  <1$ in principle are  possible here due to the 
TeV scale D-terms. However, to have such a solution in practice with  
$m_t^{({\rm pole})}\approx 175$ GeV   means having $\lambda_t (m_t)$
greater than its fixed-point value of about 1.22 with $\alpha_G 
=\alpha_s (1\, {\rm TeV})\approx 0.1 $ where the gauge couplings run 
according to the additional exotic field content as we have chosen.  

If $f A_f/g_x^2 u$, where $g_x^2 = (2/3) g_1^2 \cos^2 \alpha$ is not small, 
then Eqs.~(64) to (66) have additional contributions, but they are always 
suppressed by $v^2/u^2$ relative to $m_{12}^2 = f A_f u$, hence our 
numerical results on $\tan \beta$ and $M_{Z'}$, {\it etc.} do not change 
appreciably.  The corrections are only important if the masses and splittings 
of the two Higgs doublets are considered.

\section{Conclusions}

We have shown in this paper that there are many interesting and important 
phenomenological consequences if we assume the existence of a 
supersymmetric U(1) gauge factor at the TeV energy scale.  We assume that 
there is a Higgs superfield $S$ which is a singlet under the standard gauge 
group but which transforms nontrivially under this extra U(1) so that it 
may break the latter spontaneously without breaking the former.  We assume 
also that $H_1 H_2 S$ is an allowed term in the superpotential.  We then 
analyze the most general form of the Higgs potential and derive an upper 
limit on the lighter of the two neutral scalar Higgs bosons of the two-doublet 
Higgs sector as shown in Fig.~1.  This generalizes the well-known case of 
the Minimal Supersymmetric Standard Model (MSSM).

We then specialize to the case where this extra U(1) is derivable from a 
$E_6$ model with the particle content given by its fundamental {\bf 27} 
representation.  We discuss the effect on $Z-Z'$ mixing and the oblique 
parameters $\epsilon_{1,2,3}$, as well as the extended neutralino mass 
matrix.  We then work out in detail the consequences for supersymmetric 
scalar masses.  We note that the mere existence of a spontaneously broken 
U(1) gauge factor at the TeV scale implies new important corrections to 
these masses through the so-called D-terms which are now dominated by 
$M_{Z'}^2$ instead of just $M_Z^2$ in the MSSM.  This changes the entire 
supersymmetric scalar particle spectrum and should not be overlooked in 
future particle searches.

Assuming universal soft supersymmetry-breaking terms at the 
grand-unification (GUT) scale, we match the electroweak breaking parameters 
with the corresponding ones from the U(1) breaking.  Specifically, the 
values of $m_1^2$, $m_2^2$, and $m_{12}^2$ in the well-known two-doublet 
Higgs potential are constrained as shown by Eqs.~(64) to (69).  We then 
obtain consistent numerical solutions to these constraints and demonstrate 
how the U(1) breaking scale and the parameter $\tan \beta \equiv v_2/v_1$ 
are related through the $H_1 H_2 S$ coupling.  Our results are presented 
in Figs.~2 to 9.

During the final stage of completing this manuscript, we became aware of 
Ref. \cite{19}, which also discusses electroweak symmetry breaking with an 
additional supersymmetric U(1) gauge factor, but the emphasis there is 
on the case $f'=0$.  The case $f' \neq 0$ is also discussed there, but 
the conclusion is that whereas the breaking of the additional U(1) radiatively 
via the term $f'hh^cS$, already noted in Ref. \cite{2}, can be achieved with 
universal soft supersymmetry-breaking terms at the GUT scale, it does not work 
in the large trilinear coupling scenario.  Our approach is essentially 
orthogonal.  We concentrate on solutions where the U(1) scale is 
much larger than the electroweak scale.  With the two scales being intimately 
related through the matching of Eqs.~(64) to (66) with 
Eqs.~(67) to (69), it is in fact highly nontrivial to find solutions 
which are consistent with this matching even with an arbitrary $f'$.  
We note also that our examples are models with complete $E_6$ particle 
content and in our approximation, the Yukawa coupling $f$ is bounded 
as shown in Fig.~3.  In the more general case, the 
bound on $f$ increases as the trilinear coupling increases. 

\vspace{0.3in}
\begin{center} {ACKNOWLEDGEMENT}
\end{center}

We thank Biswarup Mukhopadhyaya for important discussions.  This work was 
supported in part by the U. S. Department of Energy under Grant No. 
DE-FG03-94ER40837.

\newpage
\bibliographystyle{unsrt}

\newpage
\begin{center} {\large \bf Figure Captions}
\end{center}

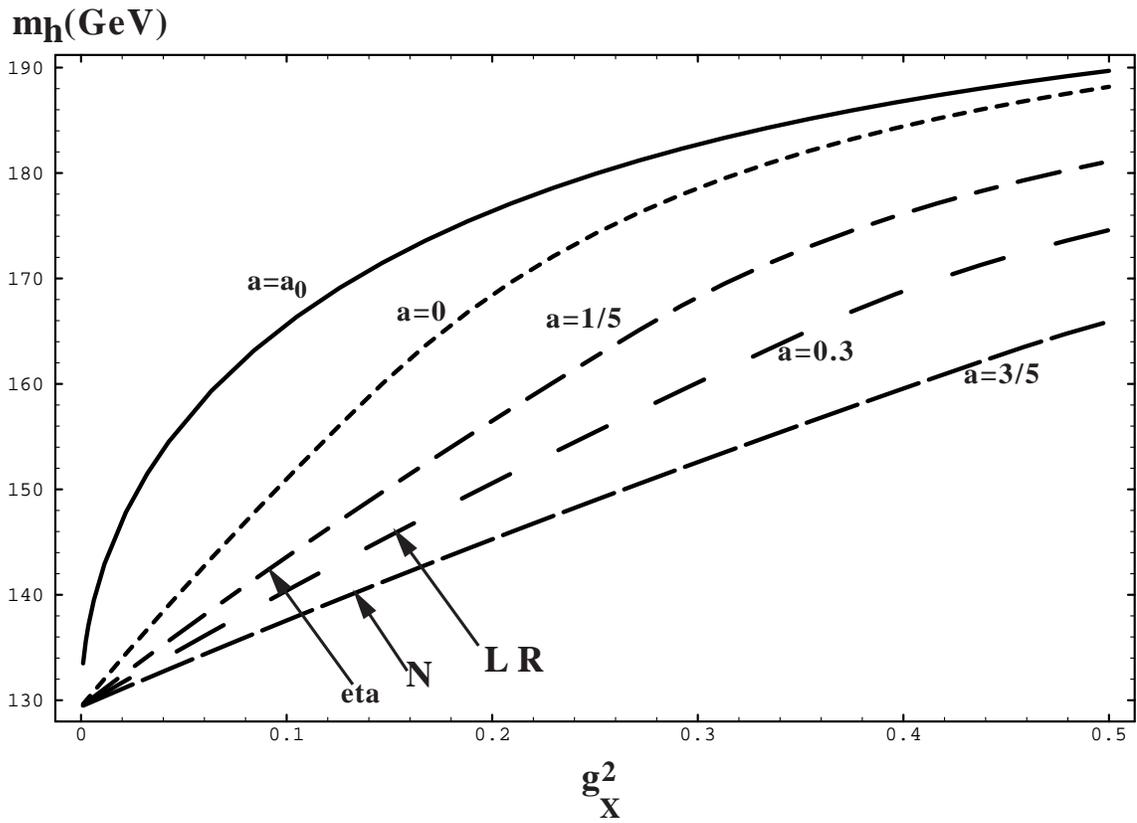
\noindent Fig.~1.  The upper bound on the lighter Higgs mass $m_h$ as a 
function of $g_X^2$ for
various values of $a$. In all cases, we find the allowed value of $f=f_0$ 
that maximizes $m_h$.
In the top curve,
we find the pair $f=f_0$ and $a=a_0$ that maximizes $m_h$ whereas the value of 
$a$ is held
fixed as labeled for the other curves. The points corresponding to the 
$\eta$, $N$, and exotic
left-right models, described in Section 3, are marked by arrows.

\noindent Fig.~2(a).  The parameter $f$ at 1 TeV as a function of $f_G=f(M_G)$ 
for models with an extra U(1) originating from $E_6$. In descending order, 
the curves represent $f'_G=f'(M_G)=0.5$, 1.0, 2.0, 3.0.	

\noindent Fig.~2(b).  The mass $M_{Z'}$ as a function of $f_G$ with the same
values $m_{\tilde g}=250$ GeV, $A_0=650$ GeV and $m_0=650$ GeV for different 
curves with the values of $f'_G$ as in 2(a).

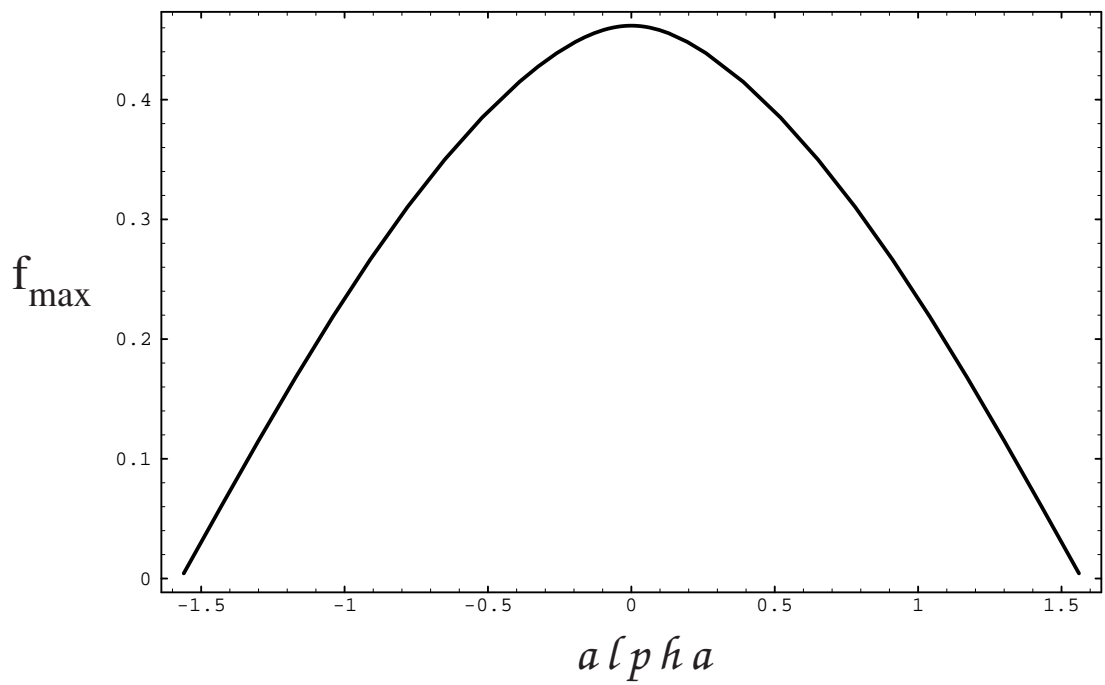
\noindent Fig.~3. The maximum value of $f=f_{\rm max}$ for which the Higgs 
potential is bounded from below as a function of $\alpha$, defined in Eq. (20).	

\noindent Fig.~4(a).  $\tan\beta$ as a function of $\alpha$ for $f=0.345$  
and $m_{\tilde g}=200$ GeV, $A_0=650$ GeV and $m_0=650$ GeV. 

\noindent Fig.~4(b).  $M_{Z'}$ (solid line), $|u|$ (short-dashed line) and
$f'|u|$ (long-dashed line) as a function of $\alpha$ for the same values 
of input parameters as in 4(a). 

\noindent Fig.~5(a).  $\tan\beta$ as a function of $\alpha$ for $f=0.345$  
and $m_{\tilde g}=300$ GeV, $A_0=950$ GeV and $m_0=950$ GeV. 

\noindent Fig.~5(b).  $M_{Z'}$ (solid line), $|u|$ (short-dashed line) and
$f'|u|$ (long-dashed line) as a function of $\alpha$ for the same
values of input parameters as in 5(a). 

\noindent Fig.~6(a).  $\tan\beta$ as a function of $f$ for $\alpha =\eta$  
and $m_{\tilde g}=250$ GeV, $A_0=650$ GeV and $m_0=650$ GeV. 

\noindent Fig.~6(b).  $M_{Z'}$ (solid line), $|u|$ (short-dashed line) and
$f'|u|$ (long-dashed line) as a function of $f$ for the same
values of input parameters as in 6(a). 

\noindent Fig.~7(a).  $\tan\beta$ as a function of $A_0$ for $\alpha =\eta$ and 
$f=0.345$, $m_{\tilde g}=250$ GeV and $m_0=650$ GeV. 

\noindent Fig.~7(b).  $M_{Z'}$ (solid line), $|u|$ (short-dashed line) and
$f'|u|$ (long-dashed line) as a function of $A_0$ for the same
values of input parameters as in 7(a). 

\noindent Fig.~8(a).  $\tan\beta$ as a function of $m_0$ for $\alpha =\eta$ 
and  $f=0.345$, $m_{\tilde g}=250$ GeV and $A_0=650$ GeV. 

\noindent Fig.~8(b).  $M_{Z'}$ (solid line), $|u|$ (short-dashed line) and
$f'|u|$ (long-dashed line) as a function of $m_0$ for the same
values of input parameters as in 8(a). 

\noindent Fig.~9(a).  $\tan\beta$ as a function of $m_{\tilde g}$ for 
$\alpha =\eta$ and $f=0.345$, $A_0=650$ GeV and $m_0=650$ GeV. 

\noindent Fig.~9(b):  $M_{Z'}$ (solid line), $|u|$ (short-dashed line) and
$f'|u|$ (long-dashed line) as a function of $m_0$ for the same 
values of input parameters as in 9(a).

\newpage
\begin{figure}[htb]
\centerline{ \DESepsf(fig1.epsf width 15 cm) } \smallskip
\nonumber
\end{figure}

\newpage
\begin{figure}[htb]
\centerline{ \DESepsf(fig2.epsf width 15 cm) } \smallskip
\nonumber
\end{figure}

\newpage
\begin{figure}[htb]
\centerline{ \DESepsf(fig3.epsf width 15 cm) } \smallskip
\nonumber
\end{figure}

\newpage
\begin{figure}[htb]
\centerline{ \DESepsf(fig4.epsf width 15 cm) } \smallskip
\nonumber
\end{figure}

\newpage
\begin{figure}[htb]
\centerline{ \DESepsf(fig5.epsf width 15 cm) } \smallskip
\nonumber
\end{figure}

\newpage
\begin{figure}[htb]
\centerline{ \DESepsf(fig6.epsf width 15 cm) } \smallskip
\nonumber
\end{figure}

\newpage
\begin{figure}[htb]
\centerline{ \DESepsf(fig7.epsf width 15 cm) } \smallskip
\nonumber
\end{figure}

\newpage
\begin{figure}[htb]
\centerline{ \DESepsf(fig8.epsf width 15 cm) } \smallskip
\nonumber
\end{figure}

\newpage
\begin{figure}[htb]
\centerline{ \DESepsf(fig9.epsf width 15 cm) } \smallskip
\nonumber
\end{figure}

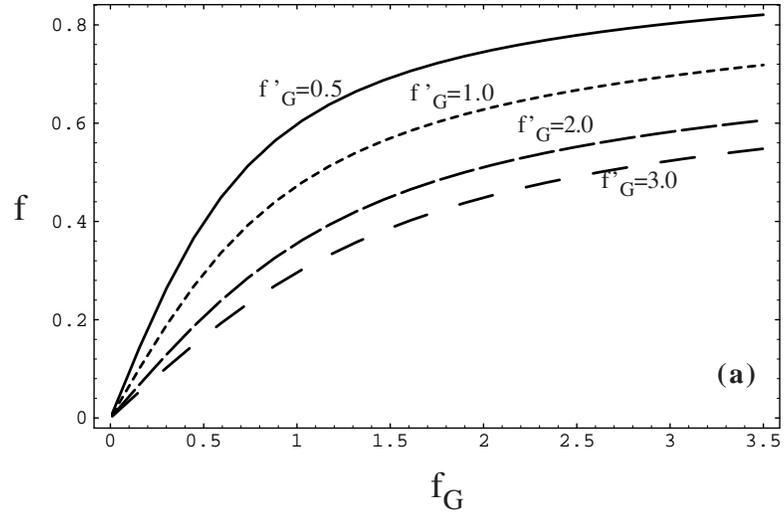
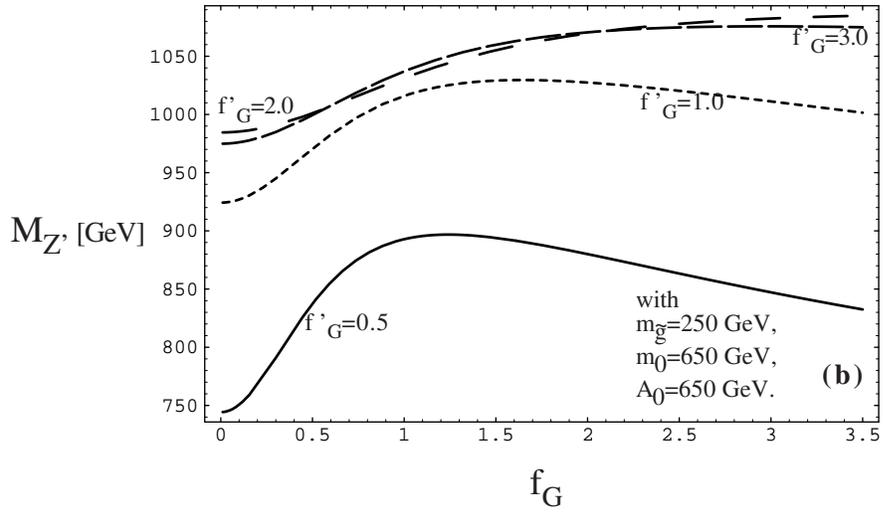
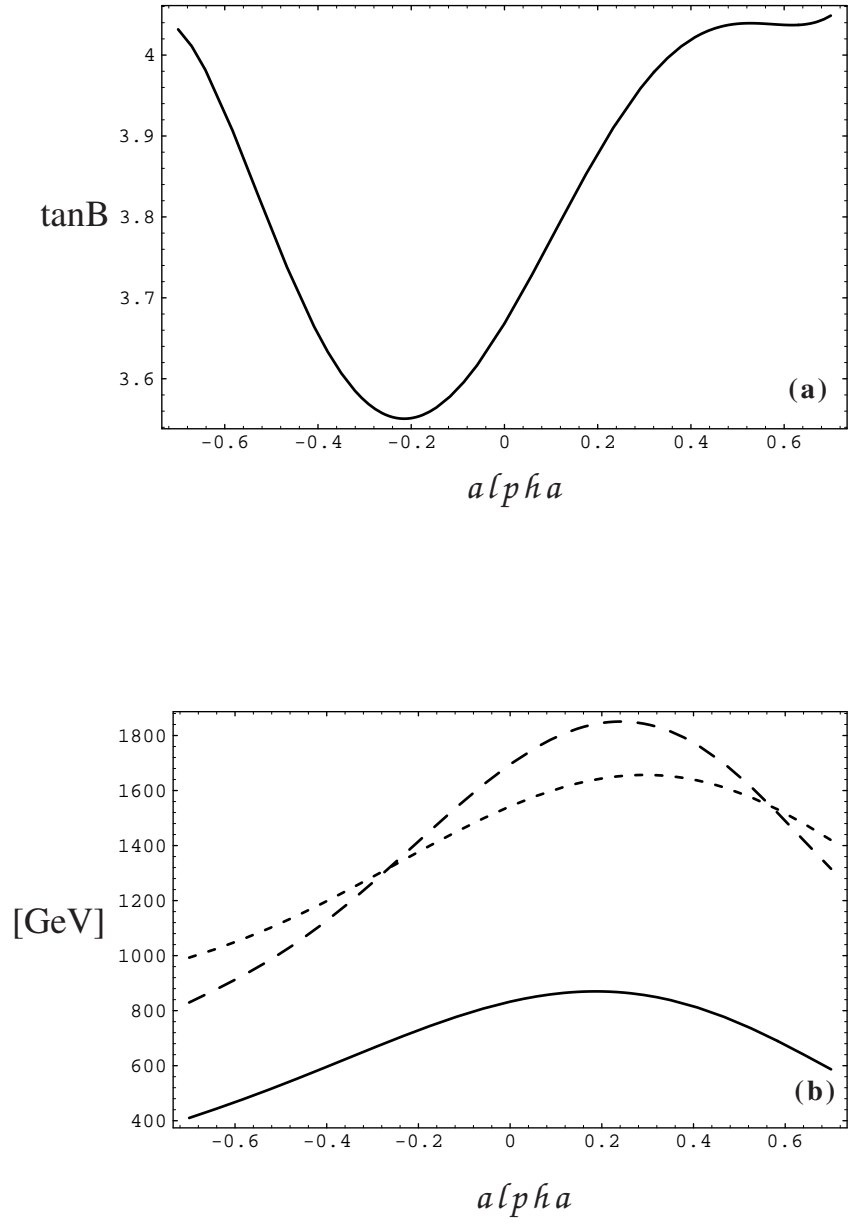
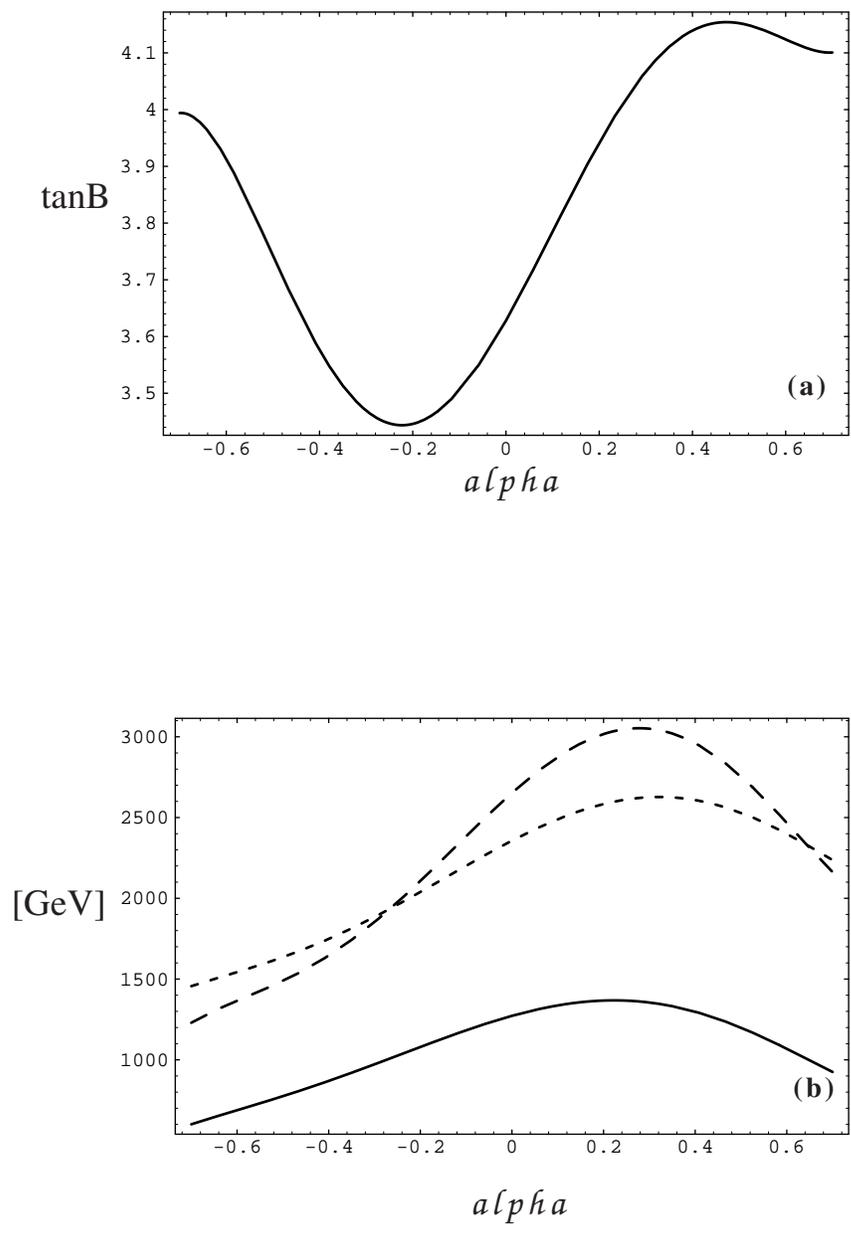
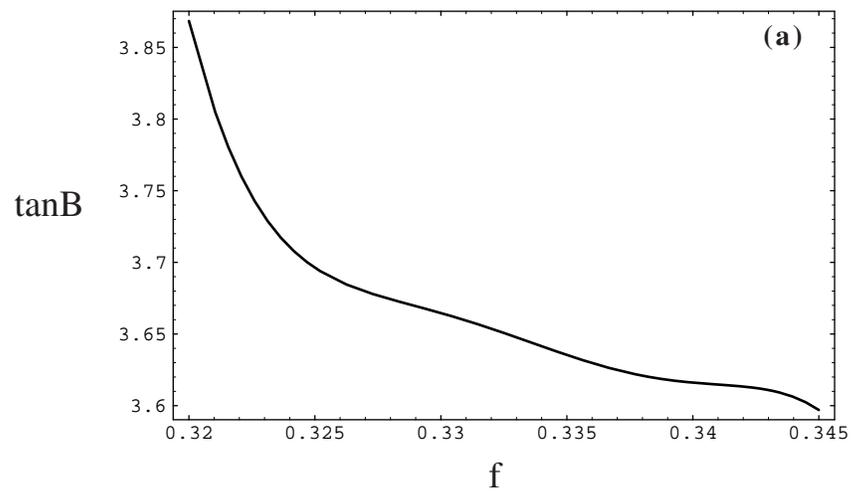
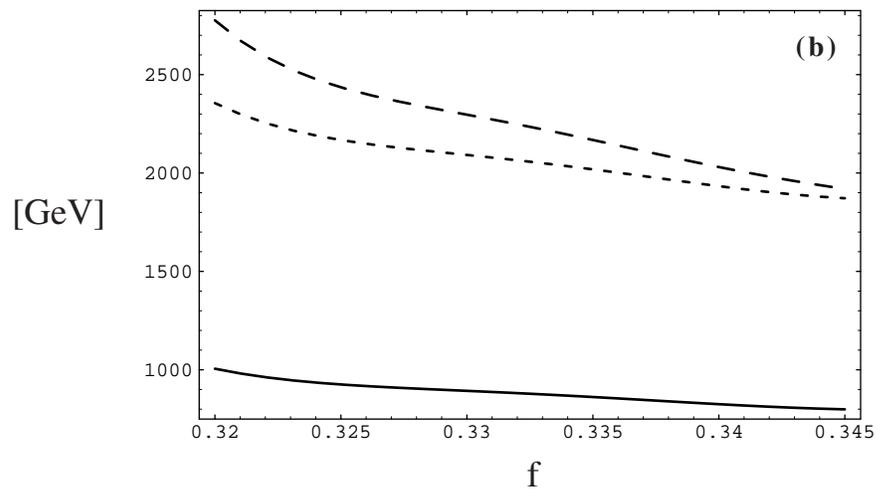
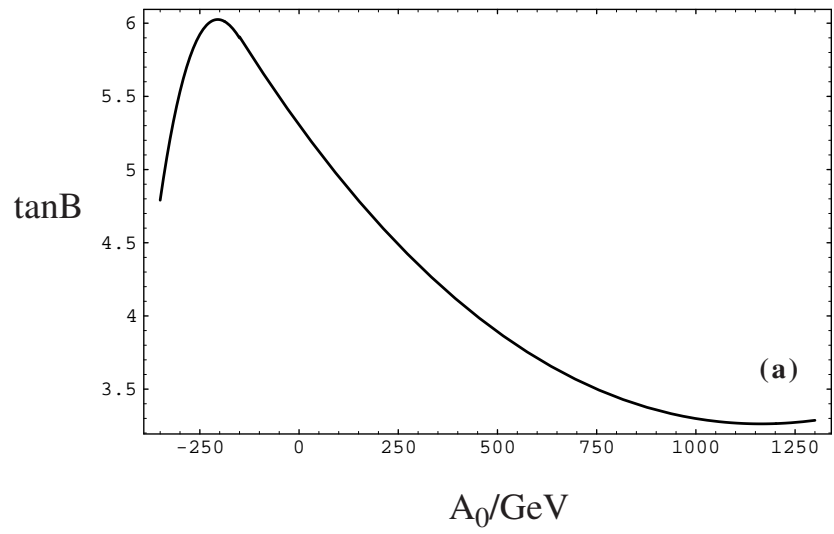
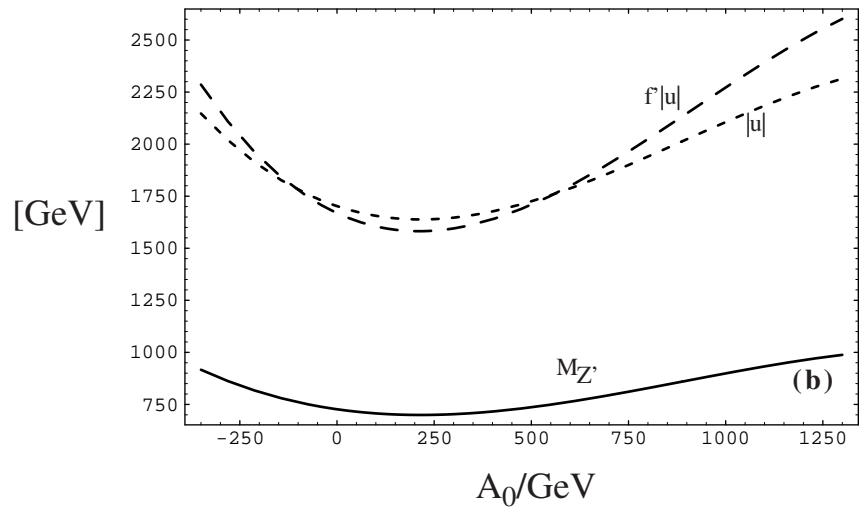
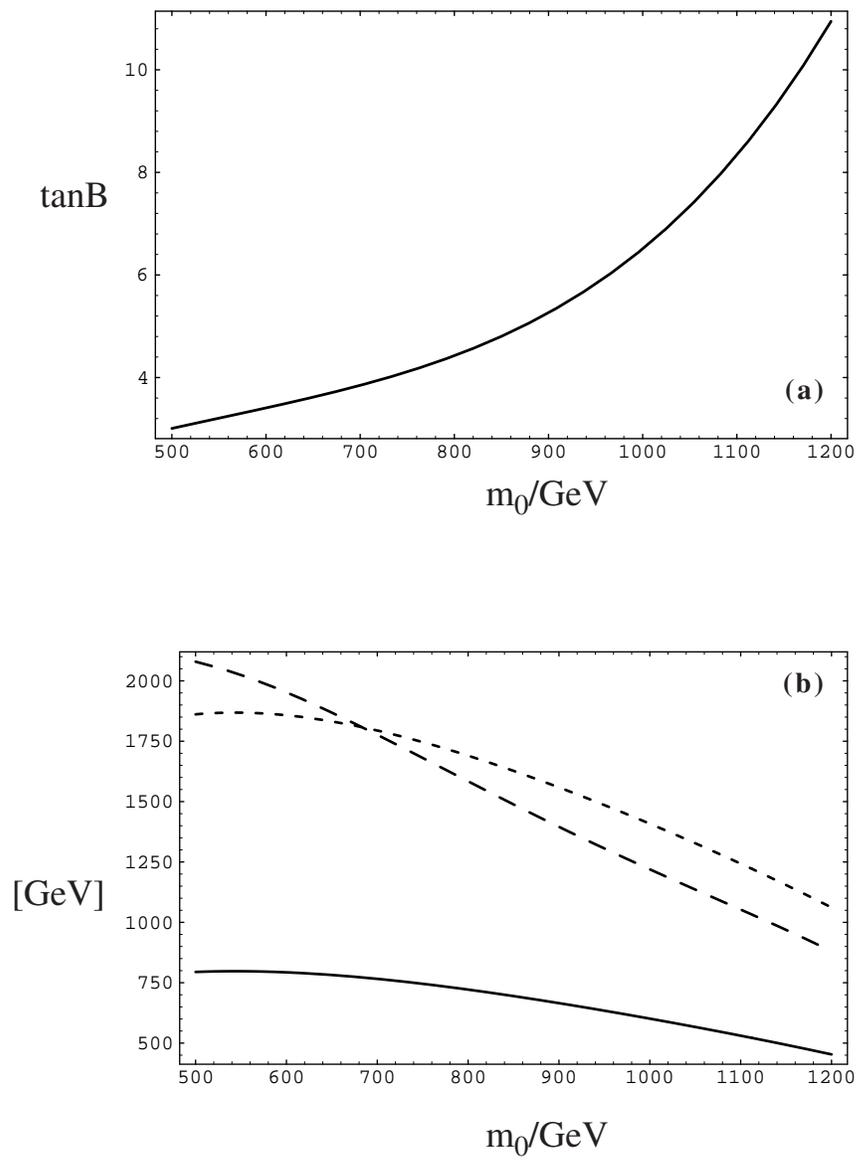
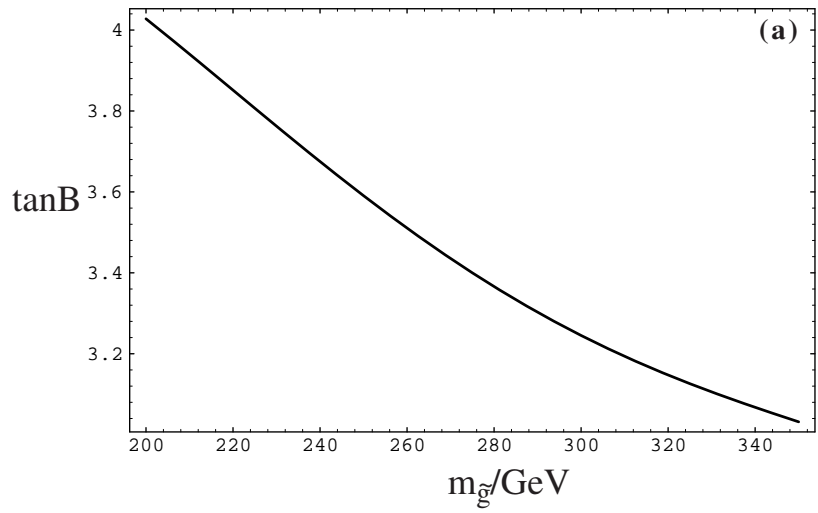
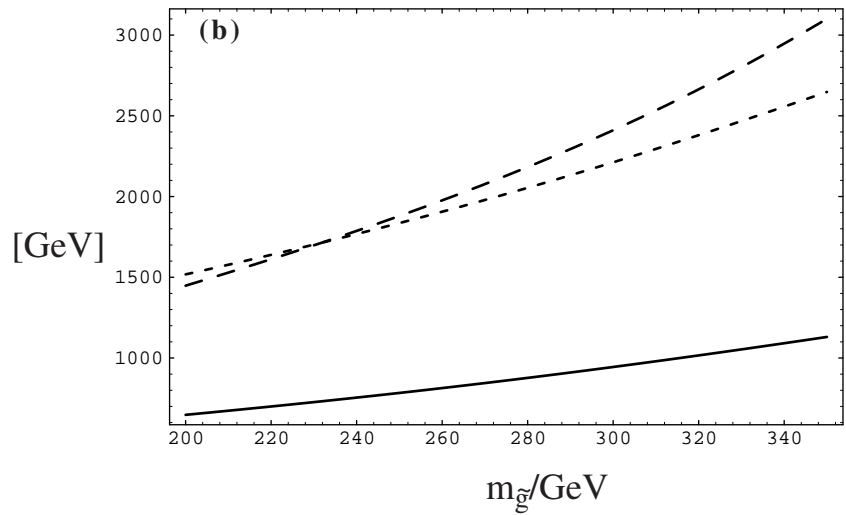
\end{document}